\documentclass{ws-ijmpd}
\usepackage[super,compress]{cite}
\usepackage[breaklinks]{hyperref}
\hypersetup{colorlinks,urlcolor=black,citecolor=black,linkcolor=black,filecolor=black}
\usepackage{breakurl}

\usepackage[]{graphicx}
\usepackage{amsmath}
\usepackage{amsfonts}
\usepackage{amssymb}
\usepackage{bm}

\usepackage{amsmath,amssymb,amsfonts,dcolumn,color,graphicx,graphics,latexsym,placeins,epsfig}

\newcommand{\be}{\begin{eqnarray}}
\newcommand{\ee}{\end{eqnarray}}
\newcommand{\ba}{\begin{eqnarray}}
\newcommand{\ea}{\end{eqnarray}}

\def\ba{\begin{eqnarray}}
\def\ea{\end{eqnarray}}
\def\be{\begin{equation}}
\def\ee{\end{equation}}

\def\ba{\begin{eqnarray}}
\def\ea{\end{eqnarray}}
\def\be{\begin{equation}}
\def\ee{\end{equation}}

\begin{document}

\thispagestyle{plain}

\def\bib{B\kern-.05em{I}\kern-.025em{B}\kern-.08em}
\def\btex{B\kern-.05em{I}\kern-.025em{B}\kern-.08em\TeX}


\markboth{Antonio Enea Romano}
{HUBBLE TROUBLE OR HUBBLE BUBBLE?}

\title{HUBBLE TROUBLE OR HUBBLE BUBBLE?}

\author{Antonio Enea Romano}

\address{Theoretical Physics Department, CERN, CH-1211 Geneva 23, Switzerland \\
Instituto de Fisica, Universidad de Antioquia, A.A.1226, Medellin, Colombia}

\maketitle




\begin{abstract}
The recent analysis of low-redshift supernovae (SN) has increased the apparent tension between the value of $H_0$ estimated from low and high redshift observations such as the cosmic microwave background (CMB) radiation.
At the same time other observations have provided  evidence of the existence of  local radial inhomogeneities  extending in different directions up to a redshift of about $0.07$. About $40\%$ of the Cepheids used for SN calibration  are directly affected because are located along the directions of these inhomogeneities. We compute with different methods the effects of these inhomogeneities on the low-redshift luminosity and angular diameter distance using an exact solution of the Einstein's equations, linear perturbation theory  and a low-redshift expansion. We confirm that at low redshift the dominant effect is the non relativist Doppler redshift correction, which is proportional to the volume averaged density contrast and to the comoving distance from the center. We derive a new simple formula relating directly the luminosity distance to the monopole of the density contrast, which does not involve any metric perturbation. We then use it to develop a new inversion method to reconstruct  the monopole of the density field from the deviations of the redshift uncorrected observed luminosity distance respect to the $\Lambda CDM$ prediction based on cosmological parameters obtained from large scale observations.

The inversion method confirms the existence of inhomogeneities whose effects were  not previously  taken into account because the $2M++$ \cite{2MPP} density field maps used to obtain the peculiar velocity  \cite{Carrick:2015xza} for redshift correction were for $z\leq 0.06$, which is not a sufficiently large scale to detect the presence of inhomogeneities extending up to $z=0.07$.
The inhomogeneity does not affect the high redshift luminosity distance because the volume averaged density contrast tends to zero asymptotically, making the value of $H_0^{CMB}$ obtained from CMB observations insensitive to any local structure. 
The inversion method can provide a unique tool to reconstruct the density field at high redshift where only SN data is available, and in particular to normalize correctly the density field respect to the average large scale density  of the Universe.
\end{abstract}


\ccode{PACS numbers: 98.80.Jk; 98.80.Es.}


\section{Introduction}
The recent analysis of low-redshift supernovae (SN) luminosity distance measurements \cite{Riess:2016jrr} have given an estimate of the the Hubble parameter  $73.24 \pm 1.74$  km s$^{-1}$ Mpc$^{-1}$ which is more than $9\%$ larger than the one obtained from  CMB data \cite{Aghanim:2016yuo}, $66.93 \pm 0.62$ km s$^{-1}$ Mpc$^{-1}$. The difference is significant at about $3.4\sigma$ confidence level, making it a discrepancy which is definitely worth investigating and  cannot be easily attributed to a statistical fluke. 
On the other hand there is  evidence \cite{Keenan:2013mfa} from luminosity density observations that  the radial density profile is not homogeneous in certain directions. In one direction, corresponding to subregion 3 in\cite{Keenan:2013mfa}, the radial profile is underdense up to about 300 Mpc/h.

In order to check the effects of local inhomogeneities on the luminosity distance and that redshift correction is sufficiently accurate we compute the full relativistic effects on the luminosity distance using an exact solution of the Einstein's equations and compare to results obtained using linear perturbations theory and a low-redshift expansion.
In the perturbative limit we obtain a formula relating directly the density field to the luminosity distance, showing that the effects are proportional to the volume average of the density field, not to the local value of the density contrast. This explains naturally why large scale observations are not affected by low redshift inhomogeneities, since their volume average is negligible on large scales.

We use this formula to derive a new inversion method to obtain the density contrast from  redshift uncorrected observed luminosity distance observation.
The method gives a value of the density contrast in subregion 3 in agreement with \cite{Keenan:2013mfa} observations.
Note that no assumption about a large void extending in all directions is made, and only the radial inhomogeneity in subregion 3 is considered, but 
since about $40\%$ of the Cepheids are affected by this inhomogeneity, the entire analysis can be affected even if the local Universe is homogeneous in other directions.


\begin{figure}[h!]
\includegraphics[width=8.5cm]{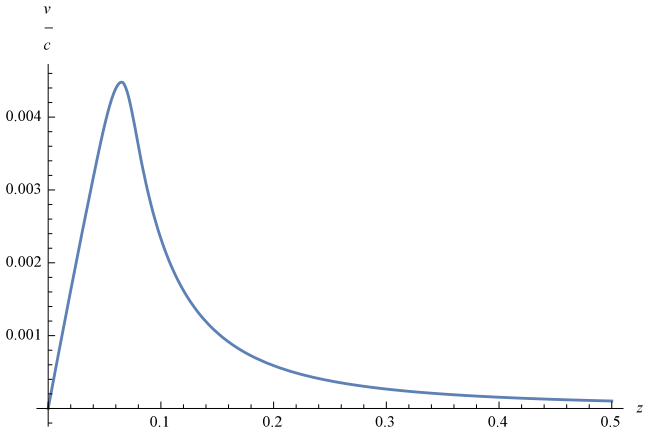}
\caption{The peculiar velocity associated to an inhomogeneity profile as the one shown in the upper panel of fig.(2)  is plotted in units of the speed of light  $c$ as a function of redshift. As can be seen the effect reaches its peak around the hedge of the inhomogeneity and is then  asymptotically suppressed due to volume averaging as shown in eq.(\ref{dzOzA}).}
\label{fig:dz}
\end{figure}


Due to the insensitivity of high redshift luminosity distance to local structure, we argue that any local deviation from the theoretical prediction of a FRW model based on cosmological parameters estimated using large scale observations such as the CMB radiation,  should be considered an evidence of local structures with a size larger than the depth of the density maps used to apply redshift correction to  observational data.

\begin{figure}[h!]
\includegraphics[width=12cm,height=7cm]{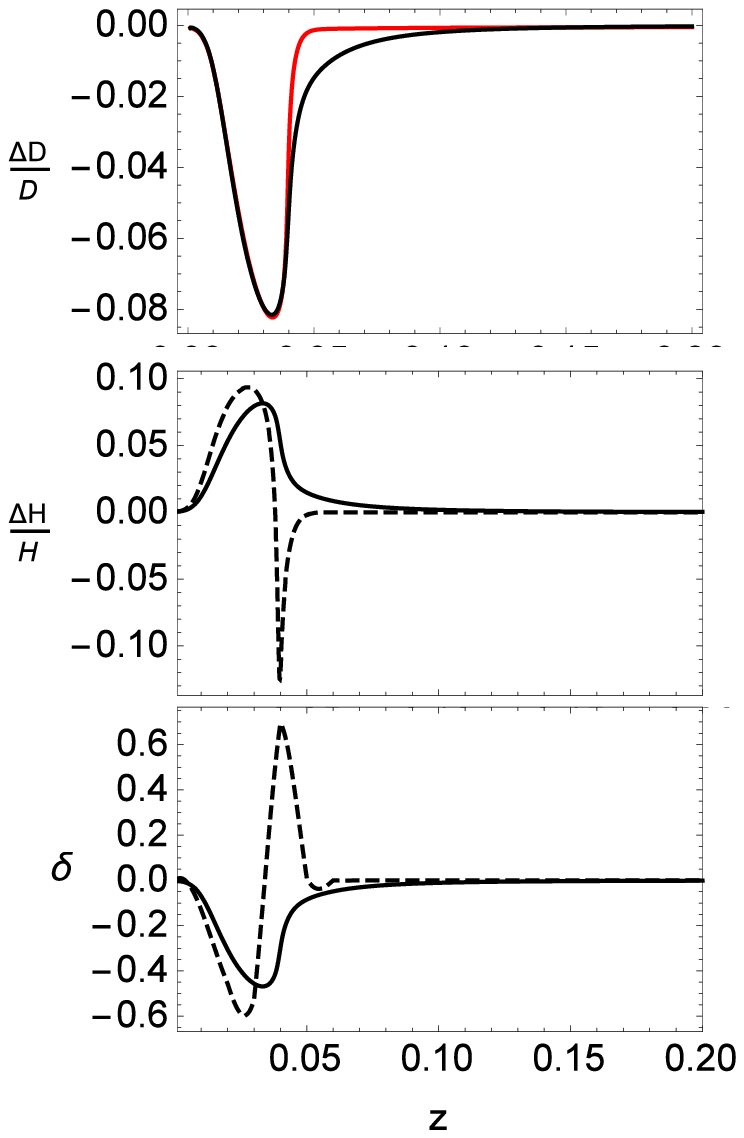}
\includegraphics[width=12cm,height=7cm]{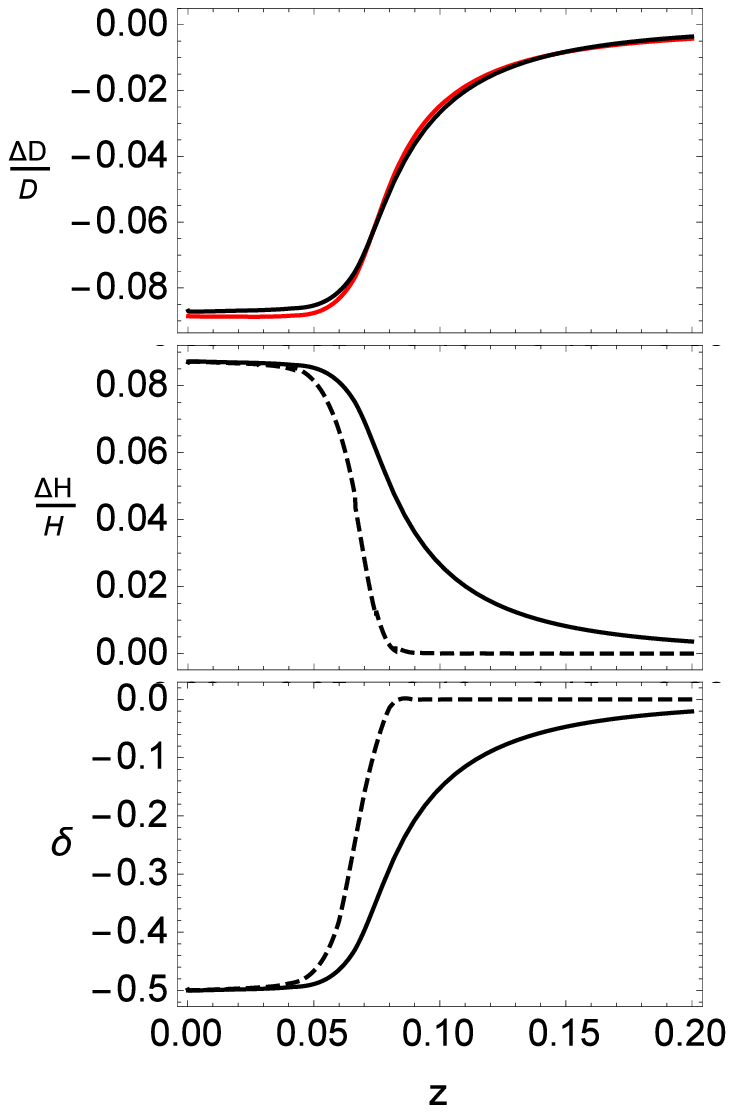}
\caption{
The  fractional difference of the luminosity distance $\Delta D_L/\overline{D_L}$ and  the local Hubble parameter $\Delta H/\overline{H_0}$  are plotted as a function of the redshift for a compensated (top) inhomogeneity such as the one studied in \cite{Romano:2014iea} and an uncompensated (bottom) void. 
 For $\Delta D/D$ the effects of the inhomogeneity are computed with a non pertubative approach using a LTB metric (red line) and are in good agreement with the approximation (black line) given in eq.(\ref{Dlowz}), confirming that  $k_v$ in eq.(\ref{kvlowz}) produces the dominant effect at low redshift. 
 The relative fractional difference $\Delta H/H$ is computed with  eq.(\ref{dHnew}) (black line) and with  eq.(\ref{H0locP}) (dashed line). For the compensated case the difference is particularly important since eq.(\ref{H0locP}) would predict a negative variation, while eq.(\ref{dHnew}) gives the correct sign, in agreement with the results of a local fitting procedure shown in fig.(\ref{fig:DH}). 
The volume averaged fractional density contrast $\overline{\delta}$ defined in eq.(\ref{dav}) is plotted with a black line and the local density contrast $\delta$ with a dashed line.
}
\label{fig:DLDHDC}
\end{figure}

The inversion method we developed could be hence particularly useful to reconstruct the density field from SN luminosity distance observations on scales where  galaxy surveys data is  not available, not just to resolve the $H_0$ tension.

\section{Effects of inhomogeneities on the luminosity distance}
Unless differently specified, we use a system of units in which 8$\pi$G=c=1.
The effects of inhomogeneities on the angular diameter  distance were calculated to first order in perturbation theory in \cite{Misao} and to second order in \cite{Barausse:2005nf}.
The results in the Newton gauge can be written in this form \cite{Bonvin:2005ps,Hui:2005nm}
\be
D_A(z)=\overline{D}_A(z) \left[1-k(z)\right] \,,\\
\ee
where $\overline{D}_A(z)$ is the diameter distance for a homogeneous Universe.
The convergence $k(z)=\sum_i k_i(z)$ is the sum of different terms among which the most important ones are
\ba
k(z)& \approx & k_{\delta}(z)+k_{v}(z) \,,\\
k_v&=&\left[1 - \frac{1}{a_e\chi_e H_e }  \right] {v_e\cdot n}+\frac{1}{a_e\chi_e H_e }v_o\cdot n  \,,\\ \label{kv}
k_{\delta}&=&  \frac{3 }{2} H_0^2 \Omega_m \int_0^{\chi_e} {\rm d} \chi
\frac{ (\chi_e - \chi)}{a_e \chi_e} \chi \delta(\chi) \,,
\ea
where $\chi$ is the comoving distance, we are denoting respectively with a lower-script $e$ or $o$ any quantity evaluated at the point of emission of the photons or at the observer, and the unit vector $n$ is in the direction of propagation from the emitter to the observer. 
The term $k_\delta$ is known as the gravitational lensing magnification \cite{1992ApJ...388..272K} while $k_v$ is due to the peculiar velocities.
The other $k_i$ terms  are related to the line of sight integral of the gravitational potential and its time derivatives, such as the integrated Sachs-Wolf effect for example, and are sub-dominant at low-redshift \cite{Bonvin:2005ps}.
We will show later using both an exact solution of the Einstein's equations and a low-redshift Taylor expansion that the most important term is $k_v$.
\section{Peculiar velocity and density maps}
In the Newtonian limit the peculiar velocity field can be related to the density field by integrating the Euler's equation \cite{peebles:1993}
\be
v(\chi)=\frac{a f H}{4 \pi}\int^{R_{Max}} \delta{(\chi')}\frac{\chi'-\chi}{|\chi'-\chi|^{3}}d^3\chi' \label{intv} \,.
\ee
In \cite{Scolnic:2013efb,Neill:2007fh,Riess:2016jrr} $R_{Max}$ corresponds to $z<0.06$, which is less than the size of the local inhomogeneity, $z\approx 0.07$. While the redshift correction method is  quite precise at low redshift as we will show later, the velocity field they obtain is missing the  peculiar velocity component due to the local void which can be obtained only by integrating eq.(\ref{intv}) over scales larger than the size of the void. 
The correct background density value $\overline{\rho}$  entering the definition of the fractional density contrast $\delta=\delta\rho/\overline{\rho}$ should be the volume average of the density $\rho$ on a scale larger than the void size,
otherwise the underdensity of the void respect to the rest of the Universe will not be taken into account.

Density maps for $z<0.06$  cover regions inside the void and cannot be used to find the relative density difference respect to the outside region, located at $z>0.07$. In other words if we only consider density maps for $z<0.06$ we cannot detect the presence of the local void, while extending the analysis to a higher redshift range \cite{Keenan:2013mfa} gives $\delta \approx-0.5$ inside the void. It turns out that the component of the peculiar velocity due to the void is crucial to explain the apparent tension in the $H_0$ estimation.

\section{Perturbative monopole correction}
The peculiar velocity associated to a spherically symmetric inhomogeneity can be obtained in the Newtonian limit as \cite{peebles:1993} 
\ba
v(\chi)&=&-\frac{1}{3}a f H \overline{\delta}(\chi){\chi} \label{vr} \,, \\
\overline{\delta}(\chi)&=&\frac{3}{4 \pi \chi^3}\int^{\chi} 4 \pi \chi'^2 \delta(\chi') \, \label{dav} d\chi'
\ea
where $\overline{\delta}$ is the density contrast averaged over the sphere of comoving radius $\chi$, it has been assumed that the density contrast can be factorized as the product of a space and a time dependent function as $\delta=A(x)D(t)$, and  $f=\frac{1}{H}\frac{\dot{D}}{D}$ is the growth factor.
The negative sign in front of eq.(\ref{vr}) implies that for an underdensity the velocity is directed outward from the center, which is what we intuitively would expect since the region outside the void is denser.
From the above equation we can see that the monopole component of the peculiar velocity is zero for a central observer, implying that  the observer velocity is not affected by the monopole component of the local structure.
Since the effects of the inhomogeneity depends on $\overline{\delta}(z)$, they extend slightly beyond the edge of the void, because the volume averaged density contrast does not go to zero immediately after it.

Eq.(\ref{vr}) is used to plot in fig.(\ref{fig:dz}) the peculiar velocity as function of the red-shift inside an inhomogeneity with a density profile given in fig.(\ref{fig:DLDHDC}). As expected the velocity is zero at the center, reaches its pick at the edge of the inhomogeneity and tends asymptotically to zero due to the volume average. 

Using  eq.(\ref{vr}), and only considering the effects due to the emitter velocity $v_e$, since, as shown in eq.(\ref{vr}), a spherical symmetric inhomogeneity does not affect the observer velocity, we can re-write $k_v$ in terms of the averaged density density and get  
\be
k_v=\frac{1}{3} f \overline{\delta} \left(a H \chi  -1 \right) \label{kvd} \,.
\ee
Note that we have used that $v_e \cdot n=\frac{1}{3} f \overline{\delta} a H \chi$ because $n$ is directed from the emitter to the observer, while in eq.(\ref{vr})  the unit vector is in the opposite direction, i.e the positive radial direction, since we are assuming a coordinate system centered at the observer.

Eq.(\ref{kvd}) is quite useful since it allows to express the effects of inhomogeneities on $D_L(z)$ directly in terms of the \textit{volume averaged density contrast} without any use of the metric.

\section{Low redshift effects of a local inhomogeneity}
At low redshift we have $a H \chi \approx z$ and consequently only the second term in eq.(\ref{kvd}) is important at leading order in $z$, giving
\be
k_v(z)=-\frac{1}{3} f\overline{\delta}(z) \label{kvlowz} \,.
\ee
In the case of $k_\delta$ we can also perform a low-redshift expansion and assuming $\overline{\delta}=\delta_c+\delta_1 z+ ..$,  the leading order term is
\be
k_\delta=\frac{3}{4} \Omega_m \delta_c  z^2\,,
\ee
which is a second order term which can be safely neglected at low redshift.

Now that we know that the dominant contribution at low redshift is $k_v$, we can compute the leading order correction to the angular diameter distance
\be
D_A(z)=D_L(z)=\overline{D_L}(z)\left[1+\frac{1}{3} f\overline{\delta}(z) \right] \,. \label{Dlowz}
\ee

In the above equation we have used that at leading order in redshift the reciprocity relation $D_L(z)=(1+z)^2D_A(z)$  implies that $D_L(z)=D_A(z)$.

As shown in fig.(\ref{fig:DLDHDC}) eq.(\ref{Dlowz}) is in very good agreement with the results obtained from the calculation of the full relativistic effects of the inhomogeneity using the LTB solution, confirming that the approximations used to derive eq.(\ref{Dlowz}) are well justified.

\section{Insensitivity of high redshift luminosity distance to the effects of local structure}
As shown in in eq.(\ref{Dlowz}) at low redshift the dominant effect of a local inhomogeneity is proportional to the volume averaged density contrast. The volume averaged density contrast of a finite size inhomogeneity tends to zero asymptotically because the total mass inside the sphere is finite while the volume keeps growing. Consequently high redshift luminosity distance observations will be unaffected by a low redshift inhomogeneity. This is important because it guarantees that any  large scale estimation of $H_0$ depending on high redshift luminosity distance, such as the $H_0$ value obtained from CMB observations for example, is insensitive to any local structure.

Expanding $k_v$ at  low-redshift  using $a H \chi \approx z$ and, as explained above $v_o=0$, we get
\ba
D_A(z)&\approx& \overline{D}_A(z)\left[1-k_v \right]\approx\overline{D}_A(z)\left[1-\frac{v_e\cdot n}{z}\right] \,.  \label{Ddz}
\ea
At low redshift we can re-write the last equation as
\ba
D_A(z) &\approx& \overline{D}_A\left(z-\delta z\right) \,, \label{Dzcorr} 
\ea
where the redshift correction is $\delta z=v_e \cdot n $, which is the dominant component of the Doppler effect in the non relativist velocity regime.
We can now get the useful relation for the redshift correction
\be
\frac{\delta z}{z} = -\frac{1}{3} f\overline{\delta}(z) \,. \label{dzOzA}
\ee

Equation (\ref{Dzcorr}) justifies the use of the  redshift correction method in the Newtonian limit, consisting in taking into account the effects of peculiar velocities on the luminosity distance by fitting the observational data with a homogeneous model after  having corrected the redshifts \cite{Neill:2007fh}. Note that for deeper and larger voids the Newtonian approach may not be sufficient, and the full relativistic calculation may be necessary.
\section{The impact of the depth of the density maps used for redshift correction}
The density maps from the 2M++ catalogue are calibrated respect to them self, i.e. the average density is assumed to be the average density of the catalogue. If the region surveyed is embedded in an inhomogeneity whose size is larger then the catalogue depth, the density contrast with respect to the average density of the Universe is miss-estimated. In this case in fact the total density contrast $\delta$ has two components
\ba
\delta = \delta_{loc}+\delta_{LS}
\ea
where we are denoting with $\delta_{loc}$ the local component which can be probed by 2M++ and with  $\delta_{LS}$ the  large scale component which 2M++ cannot detect. For an observer which has only access to data up to the depth of 2M++ there is in fact  no way to establish the relation between the average density of the Universe and the average density inside 2M++. 

This also implies that the peculiar velocity field \cite{Carrick:2015xza} obtained by integrating the density field using eq.(\ref{intv}) will have two components, 
\ba
v &=& v_{loc}+v_{LS} \,, \\
v_{loc} &=& \frac{a f H}{4 \pi}\int^{R_{Max}} \delta_{loc}{(\chi')}\frac{\chi'-\chi}{|\chi'-\chi|^{3}}d^3\chi' \label{intv} \,, \\
v_{LS} &=& \frac{a f H}{4 \pi}\int^{R_{Max}} \delta_{LS}{(\chi')}\frac{\chi'-\chi}{|\chi'-\chi|^{3}}d^3\chi' \label{intv} \,. 
\ea
For inhomogeneities such has the one in subregion 3 the redshift correction applied in \cite{Riess:2016jrr} is only taking into account $v_{loc}$ and neglecting the effect due to $v_{LS}$ can induce a miss-estimation of the luminosity distance.
\section{Reconstructing the density field from the luminosity distance}
Since the luminosity distance is affected by inhomogeneities we can use it to infer the density profile which is causing it to deviate from the one of a homogeneous Universe. 
We can use  eq.(\ref{kvd}) to obtain the monopole component of the averaged density contrast from the angular diameter distance 
\be
\overline{\delta}=\frac{3}{f}\left(1-\frac{D_A}{\overline{D}_A}\right)\frac{1}{(a H \chi-1)} \,,
\ee
After differentiating once $\overline{\delta}$ respect to $\chi$ we can also get the density contrast as a function of the comoving distance
\be
\delta (\chi )= \frac{1}{3}(\chi \overline{\delta}'   +3 \overline{\delta})\approx \frac{1}{3}\left[\chi \frac{d\overline{\delta}}{dz}\left(\frac{d\chi}{dz}\right)^{-1}   +3 \overline{\delta}\right]\,, \label{dinv}
\ee
where we are denoting with $\overline{\delta}' $ the derivative of $\overline{\delta}$ respect to $\chi$.
For the case of a constant  $\overline{\delta}=\delta_c$ the formula gives the expected result $\delta(\chi)=\delta_c$. 
The formula above represents the main theoretical result of this paper, and it allows to reconstruct directly the density field from luminosity distance observations without determining the metric as required with other much more complicated inversion methods \cite{Inversion,Chung:2006xh,Tokutake:2016hod}, and it  has the additional the advantage of being valid also in presence of cosmological constant.
Using this inversion method new evidence of local structure could be obtained from luminosity distance data, which can be then further investigated using other type of observations such as luminosity density or number counts for example.

Assuming $f={\Omega_m}^{0.55}$ \cite{peebles:1993} and   $D_A/{\overline{D}_A}\approx\Delta H/\overline{H_0} \approx 9.4$ we get a value of the volume averaged density contrast $\overline{\delta}\approx -0.57$, in the range which has been estimated for subregion 3 in \cite{Keenan:2013mfa}, confirming the existence of the inhomogeneity found in that direction. 
In the estimation of $H_0$ performed in \cite{Riess:2016jrr} about $20\%$ of SN and $40\%$ of the Cepheids used for calibration  are in the subregions 2 and 3 defined in \cite{Keenan:2013mfa}. 
The inhomogeneities detected in those directions can  consequently have some important  impact on estimating the Hubble constant, even if the other directions are homogeneous.

\section{The effects on  the estimation of $H_0$}
The analysis performed in \cite{Keenan:2013mfa} was not covering the full sky, and inhomogeneities were detected only in some directions. Here we will assume that shear effects are negligible at low redshift  and that the effects in a given direction can be approximated as the effects of a spherically symmetric inhomogeneity with the same radial profile as the one detected in that particular direction.
We are not assuming the existence of a unique spherically symmetric inhomogeneity centered around us to model local structure, but using a spherical approximation as a tool to compute the effects of the inhomogeneity profiles detected in different directions in \cite{Keenan:2013mfa}. In other words we are respecting the observed anisotropy of local structure by choosing different radial profiles in different directions, and there is no fine tuning in choosing the position of the observer.
The effects we compute should not be compared to the full sky Hubble diagram but to the directional Hubble diagram for SN in the specific direction along which the density profile is known. 

The local estimation of $H_0$ is based on the implicit assumption that the Universe is homogeneous and isotropic and can  consequently be modeled with a FRW metric, which corresponds to this low-redshift expansion for the luminosity distance
\begin{eqnarray}
D_L(z)&=&\frac{1}{H_0} z+..\,,
\end{eqnarray}
from which at low redshift we can define
\ba
H_0^{loc}&=& \lim_{z \to 0} \frac{z}{D_L(z)} \,. \label{H0loc}
\ea
If the background luminosity distance is modified by perturbations we have
\ba
D_L(z)&=&\overline{D_L}(z)+\Delta D_L(z) 
\ea
and at leading order, after combining  eq.(\ref{H0loc}) and (\ref{Dlowz}), we get 
\ba
H_0^{loc}&=&\overline{H_0}+\Delta H(z) \quad, \quad \overline{H_0}= \lim_{z \to 0} \frac{z}{\overline{D}_L(z)} \\
\frac{\Delta H(z)}{\overline{H_0}}&=&-\frac{\Delta D_L(z)}{\overline{D}_L(z)}=-\frac{1}{3}f\overline{\delta}(z) \label{dHnew}
\ea

The local variation of the Hubble parameter is proportional to the volume average of the density contrast $\overline{\delta}$ while previous \cite{Turner,peebles:1993} calculations, based on the Newtonian limit of cosmological scalar perturbation theory applied to the dynamical definition of the Hubble parameter $H=\dot{a}/{a}$,  were giving
\ba
\frac{\Delta H}{\overline{H}}=-\frac{1}{3}f\delta \,. \label{H0locP}
\ea
This difference is particular important when considering the effects of local inhomogeneities which do not have constant density contrast such as compensated voids for example or near the edge of the inhomogeneity. For example even if the local density $\delta(z)$ is zero the volume averaged $\overline{\delta}(z)$ could be different from zero if there is an inhomogeneity at lower redshift as shown in fig.(\ref{fig:DLDHDC}).
Another important difference is that eq.(\ref{H0locP}) cannot be used to obtain any information about the large scale value of $H_0$, because  by construction it is only valid locally, while eq.(\ref{dHnew}) correctly predicts that the  high redshift effect of a local inhomogeneity tends to zero because the volume averaged  fractional density contrast $\overline{\delta}$ is asymptotically negligible.


We show in fig.(\ref{fig:DLDHDC}) the plot of $\overline{\delta}(z)$ for a local void of the type supported by luminosity density observations \cite{Keenan:2013mfa}.
The formula in eq.(\ref{kv}) for the computation of the effects of inhomogeneities on the luminosity distance is based on perturbation theory and its precision is limited by the validity of the assumption that the inhomogeneity can indeed be modeled with a pertubative approach. For the inhomogeneities we are considering the results of the pertubative computation are in good agreement with the exact results as shown in fig.(\ref{fig:DLDHDC}), confirming that $k_v$ is the dominant term. 
The exact calculation is performed using a Lema\'itre-Tolman-Bondi (LTB) metric, which is a dust spherically symmetric solution,  with density contrast profiles given in fig.(\ref{fig:DLDHDC}), based on the observational evidence of the inhomogeneities found in different directions in \cite{Keenan:2013mfa} and previous studies \cite{Romano:2014iea}.  

\begin{figure}[h!]
\includegraphics[width=8.5cm,height=3.5cm]{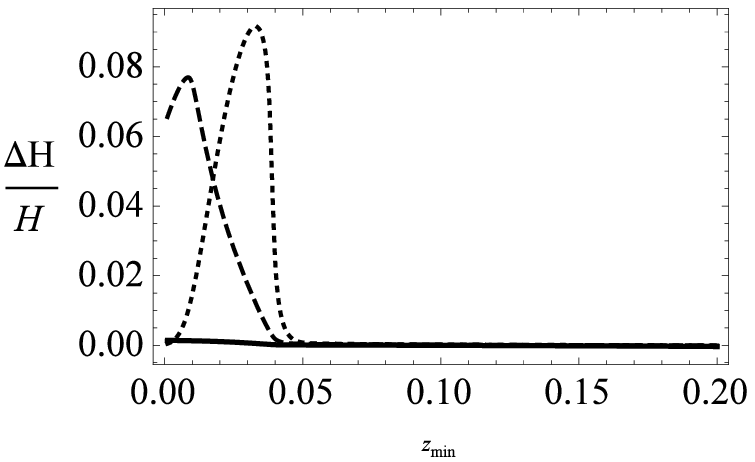}
\includegraphics[width=8.5cm,height=3.5cm]{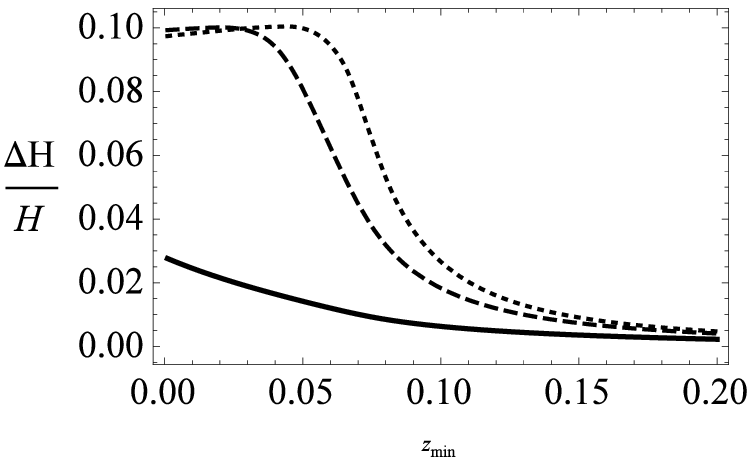}

\caption{
The fractional  difference respect to the background  $\Delta H/H$, estimated according to the same procedure used for  fig.(12) in \cite{Riess:2016jrr}  by fitting $H_0$ from luminosity distance data in a range $z_{min}<z<z_{min}+\Delta z$, is plotted as a function of $z_{min}$. The black line is for $\Delta z=0.15$ as in \cite{Riess:2016jrr}, the dashed for $\Delta z=0.03$ and the dotted line for $\Delta z=0.0015$. The luminosity distance used as input for the $H_0$ fit is the one for the compensated (top) and uncompensated (bottom) inhomogeneities shown in fig.(\ref{fig:DLDHDC}). For $\Delta z=0.15$ the fitting interval is much larger than the size of the inhomogeneity and consequently the fitted $H_0$ is affected at less than $\approx 0.1\% $ level for a compensated inhomogeneity because the effects of the homogeneity are smeared out, while for an uncompensated inhomogeneity the effect can be up to $\approx 2\% $. For $\Delta z=0.03$ and $\Delta z=0.0015$ the effect is clearly noticeable for both a compensated and an uncompensated inhomogeneity. This shows that, even if a compensated inhomogeneity were present, a fitting procedure such as the one used  for  fig.(12) in \cite{Riess:2016jrr} would not be able to detect it unless a sufficiently small $\Delta z$ were used. Furthermore it should be noted that if the inhomogeneity is extending only in some direction the effect on the full sky Hubble diagram would be much smaller than in the above plots, since only some SN would affected. Consequently the above plots are just given as an example of the directional effects and should  be compared to the directional Hubble diagram obtained only from SN in a given direction, not to the full sky diagram. 
}
\label{fig:DH}
\end{figure}

\section{Comparison with previous results}
The effects of inhomogeneities on the luminosity distance have been an active research field for quite some time and it is important to emphasize what are the new results obtained in this paper.

The main theoretical new results are in eq.(\ref{Dlowz}), eq.(\ref{dzOzA}) and eq.(\ref{dHnew}). These formulae show that the effects of inhomogeneities depend on the volume average of the density contrast $\overline{\delta}(z)$, not on its local value $\delta(z)$, contrary to what it was previously assumed \cite{Turner}. The comparison with exact general relativistic calculations, as shown in fig.(\ref{fig:DLDHDC}), shows that the perturbative approximation  we have used is enough to estimate the effects of a local void of the size necessary to resolve the apparent $H_0$ discrepancy.
This result is agreement with Gauss's law, and is particularly important to understand the effects of inhomogeneities with a non constant density profile as shown in fig.(\ref{fig:DLDHDC}). The other important consequence if that this formula can simply explain why high redshift luminosity distance observations are not affected by a local inhomogeneity, since the volume average in this case tends to zero because the volumes keeps growing.
Note  for example that using eq.(\ref{H0locP}) in the case of a compensated void we would get the wrong result as function of the redshift, while with eq.(\ref{dHnew}) we get that the overdensity and the underdensity can compensate each other producing an effect with opposite sign with respect to the one obtained with eq.(\ref{H0locP}), as shown in the upper panel of fig.(\ref{fig:DLDHDC}).  

The other important  theoretical result is the new inversion method derived in eq.(\ref{dinv}), which allows to reconstruct the monopole of the density contrast from luminosity distance observation. Previous inversion methods required the solution of complicated systems of differential equations \cite{Chung:2006xh,Inversion,Tokutake:2016hod}, while the one derived in this paper is much simpler, and quite accurate as shown in fig.(\ref{fig:DLDHDC}).

From the point of view of experimental data analysis we have pointed out that red-shift correction performed by \cite{Riess:2016jrr} may be biased by the depth of the $2M++$ survey, which could be not enough to detect the presence of a inhomogeneity extending beyond $z=0.06$. We have also observed that there is a high concentration (about $40\%$) of SN and Cepheids in directions in which inhomogeneities have been detected \cite{Keenan:2013mfa} using luminosity density data, and this could make these effects important even if the void does not extend in all directions. In this case it would not be a spherical void, but an anisotropic inhomogeneity with different density profiles in different directions, but the effect would still be statistically relevant if a large number of data points are in inhomogeneous direction, as observed in \cite{Romano:2014iea}.

\section{Conclusions}
We have studied the effects produced  by a  local inhomogeneities on the luminosity distance. The effects have been computed using different methods which all agree at low redshift, confirming the redshift correction is the dominant effect. A simple formula relating the volume average of the density contrast to the luminosity distance has been derived. A new inversion method to reconstruct the monopole of the density profile from luminosity distance observations has been developed, and has confirmed the existence of inhomogeneities previously detected using luminosity density observations in \cite{Keenan:2013mfa}. These inhomogeneities can have an important impact on the estimation of $H_0$ because  about  $40\%$ of the Cepheids used for SN calibration are affected. 
These effects where not taken into account in previous analysis  \cite{Riess:2016jrr} because the density maps used to find the redshift corrections due to peculiar flows were not being integrated on  sufficiently large scales due to the limited depth of $2M++$ as compared to the size of these inhomogeneities.

The radial density profile inferred from luminosity density observations \cite{Keenan:2013mfa} is not the same in all directions and this could induce some additional direction dependent corrections to the estimation of the Hubble constant as noted in \cite{Romano:2014iea}. 
In the future it will be interesting to develop  a new inversion method to reconstruct the density field  in different directions in order to obtain information about large scale structure on scales which cannot be probed by galaxy catalogues. This will indeed provide a unique method to study the density field at high redshift where only SN data is available. 

\appendix
\section{Local relation between $\Delta H$ and $\delta$}
Let us consider a small perturbation of the scale factor parametrized as 
\ba 
a&=&a_b(1-\epsilon)=a_b+\delta a \,.
\ea
In the non relativist limit, i.e. neglecting the pressure,  the continuity equation for the background can be written as $d(\rho a^3)=d\rho \, a^3+3\rho \, a^2 da=0$,
from which we can get
\ba
\delta&=&\frac{\delta\rho}{\rho}=-3\frac{\delta a}{a}=3\epsilon \,.
\ea
 We can now compute the perturbed Hubble parameter 
 \ba
 H&=&\frac{\dot{a}}{a}=\frac{\dot{a_b}}{a_b}-\dot{\epsilon}=H_b-\frac{1}{3}\dot{\delta}=H_b+\Delta H \,.
 \ea
 where 
 \ba
 H_b&=&\frac{\dot{a_b}}{a_b} \\
 \Delta H&=&-\frac{1}{3}\dot{\delta} \,.
 \ea
 
If we assume \cite{peebles:1993} the density contrast can be written as $\delta(t)=A(x)D(t)$ we get
\ba
\frac{\Delta H}{H_b}&=&-\frac{1}{3}f \delta \label{dHHb} \,.\\
f&=&\frac{1}{H_b}\frac{\dot{D}}{D} \,.
\ea
This relation is based on scalar perturbation theory and is only valid in the non relativistic limit, i.e. when pressure effects are negligible.
Since all quantities in the above equations are local perturbations, eq.(\ref{dHHb}) is only valid at a given point in space time and is not directly related to the value of $H_0$ which is obtained from SN data, defined in terms of the luminosity distance, and not in terms of the scale factor $a(t)$, which is not directly observable. The correct formula to estimate the effects  of local structure on the $H_0$ value obtained from luminosity distance observations is given in eq.(\ref{dHnew}).

As shown above eq.(\ref{dHHb}) is based on the non the non relativist limit of scalar perturbation theory and it does not involve any  spatial or statistical averaging procedure.  Nevertheless it has been applied to estimate the variance of $H_0$ \cite{Turner} due to local structure, interpreting perturbations not as local quantities but as the standard deviation obtained from some statistical average. In the future it will be interesting to study the non pertubative effects, which could be important also at low redshift,as shown for example for the relation between the metric and number counts  \cite{Vallejo:2017rga}.

This approach is not well justified in general relativity, since spatial averaging may ignore important effects from back-reaction \cite{Buchert:1999er}.
On the contrary the formula in eq.(\ref{dHnew}) is only based on the observational definition of $H_0$, and it does no require any averaging procedure. The spatial average of the density contrast appearing in eq.(\ref{dHnew}) comes from the integration of the Euler's equations and is reminiscent of  the Gauss's law in Newtonian gravity, since it is obtained from the non relativist limit of linear cosmological perturbations theory.

~\\{\bf Acknowledgments:}\\
I thank Misao Sasaki, George P. Efstathiou, Adam Riess and Malcolm Fairbairn for useful discussions and comments.

\bibliographystyle{ws-ijmpd}
\bibliography{bib}

\begin{thebibliography}{10}

\bibitem{2MPP}
G.~{Lavaux} and M.~J. {Hudson}, {\em Mon. Not. Roy. Astron. Soc.} {\bf 416}
  (October 2011) 2840, \href{http://arxiv.org/abs/1105.6107}{{\ttfamily
  arXiv:1105.6107}}.

\bibitem{Carrick:2015xza}
J.~Carrick, S.~J. Turnbull, G.~Lavaux and M.~J. Hudson, {\em Mon. Not. Roy.
  Astron. Soc.} {\bf 450}  (2015) 317,
  \href{http://arxiv.org/abs/1504.04627}{{\ttfamily arXiv:1504.04627
  [astro-ph.CO]}}.

\bibitem{Riess:2016jrr}
A.~G. Riess {\em et~al.}, {\em Astrophys. J.} {\bf 826}  (2016)  ~56,
  \href{http://arxiv.org/abs/1604.01424}{{\ttfamily arXiv:1604.01424
  [astro-ph.CO]}}.

\bibitem{Aghanim:2016yuo}
 Planck Collaboration (N.~Aghanim {\em et~al.})  (2016)
  \href{http://arxiv.org/abs/1605.02985}{{\ttfamily arXiv:1605.02985
  [astro-ph.CO]}}.

\bibitem{Keenan:2013mfa}
R.~C. Keenan, A.~J. Barger and L.~L. Cowie, {\em Astrophys. J.} {\bf 775}
  (2013)  ~62, \href{http://arxiv.org/abs/1304.2884}{{\ttfamily arXiv:1304.2884
  [astro-ph.CO]}}.

\bibitem{Romano:2014iea}
A.~E. Romano and S.~A. Vallejo, {\em Europhys. Lett.} {\bf 109}  (2015)
  39002, \href{http://arxiv.org/abs/1403.2034}{{\ttfamily arXiv:1403.2034
  [astro-ph.CO]}}.

\bibitem{Misao}
M.~{Sasaki}, {\em Mon. Not. Roy. Astron. Soc.} {\bf 228} (October 1987) 653.

\bibitem{Barausse:2005nf}
E.~Barausse, S.~Matarrese and A.~Riotto, {\em Phys. Rev.} {\bf D71}  (2005)
  063537, \href{http://arxiv.org/abs/astro-ph/0501152}{{\ttfamily
  arXiv:astro-ph/0501152 [astro-ph]}}.

\bibitem{Bonvin:2005ps}
C.~Bonvin, R.~Durrer and M.~A. Gasparini, {\em Phys. Rev.} {\bf D73}  (2006)
  023523, \href{http://arxiv.org/abs/astro-ph/0511183}{{\ttfamily
  arXiv:astro-ph/0511183 [astro-ph]}}, [Erratum: Phys. Rev.D85,029901(2012)].

\bibitem{Hui:2005nm}
L.~Hui and P.~B. Greene, {\em Phys. Rev.} {\bf D73}  (2006)   123526,
  \href{http://arxiv.org/abs/astro-ph/0512159}{{\ttfamily
  arXiv:astro-ph/0512159 [astro-ph]}}.

\bibitem{1992ApJ...388..272K}
N.~{Kaiser}, {\em APJ} {\bf 388} (April 1992) 272.

\bibitem{peebles:1993}
P.~J.~E. Peebles, {\em {Principles of physical cosmology}} (Princeton
  University Press, 1993).

\bibitem{Scolnic:2013efb}
D.~Scolnic {\em et~al.}, {\em Astrophys. J.} {\bf 795}  (2014)  ~45,
  \href{http://arxiv.org/abs/1310.3824}{{\ttfamily arXiv:1310.3824
  [astro-ph.CO]}}.

\bibitem{Neill:2007fh}
 SNLS Collaboration (J.~D. Neill, M.~J. Hudson and A.~J. Conley), {\em
  Astrophys. J.} {\bf 661}  (2007)   L123,
  \href{http://arxiv.org/abs/0704.1654}{{\ttfamily arXiv:0704.1654
  [astro-ph]}}.

\bibitem{Inversion}
A.~E. Romano, H.-W. Chiang and P.~Chen, {\em Class. Quant. Grav.} {\bf 31}
  (2014)   115008, \href{http://arxiv.org/abs/1312.4458}{{\ttfamily
  arXiv:1312.4458 [astro-ph.CO]}}.

\bibitem{Chung:2006xh}
D.~J.~H. Chung and A.~E. Romano, {\em Phys. Rev.} {\bf D74}  (2006)   103507,
  \href{http://arxiv.org/abs/astro-ph/0608403}{{\ttfamily
  arXiv:astro-ph/0608403 [astro-ph]}}.

\bibitem{Tokutake:2016hod}
M.~Tokutake and C.-M. Yoo, {\em JCAP} {\bf 1610}  (2016)   009,
  \href{http://arxiv.org/abs/1603.07837}{{\ttfamily arXiv:1603.07837
  [astro-ph.CO]}}.

\bibitem{Turner}
E.~L. {Turner}, R.~{Cen} and J.~P. {Ostriker}, {\em ApJ} {\bf 103} (May 1992)
  1427.

\bibitem{Vallejo:2017rga}
S.~A. Vallejo and A.~E. Romano, {\em JCAP} {\bf 1710}  (2017)   023,
  \href{http://arxiv.org/abs/1703.08895}{{\ttfamily arXiv:1703.08895
  [astro-ph.CO]}}.

\bibitem{Buchert:1999er}
T.~Buchert, {\em Gen. Rel. Grav.} {\bf 32}  (2000) 105,
  \href{http://arxiv.org/abs/gr-qc/9906015}{{\ttfamily arXiv:gr-qc/9906015
  [gr-qc]}}.

\end{thebibliography}
\end{document}